\documentclass[aip,reprint]{revtex4-1}
\usepackage{graphicx}
\usepackage{amsmath}

\draft 

\begin{document}


\title{Mid-Infrared Hyperspectral Microscopy with Broadband 1-GHz Dual Frequency Combs} 



\author{Peter Chang}
\email{peter.chang-1@colorado.edu}
 \affiliation{Department of Electrical, Computer and Energy Engineering, University of Colorado at Boulder, CO U.S.A.}
 \affiliation{Department of Physics, University of Colorado at Boulder, CO U.S.A.}

\author{Ragib Ishrak}%
\affiliation{Department of Electrical and Computer Engineering, University of Houston, TX U.S.A.}

\author{Nazanin Hoghooghi}
\affiliation{Department of Mechanical Engineering, University of Colorado at Boulder, CO U.S.A.}
\affiliation{Time and Frequency Division, National Institute of Standards and Technology, Boulder, CO U.S.A.}

\author{Scott Egbert}
\affiliation{Department of Mechanical Engineering, University of Colorado at Boulder, CO U.S.A.}

\author{Daniel Lesko}
\affiliation{Department of Chemistry, University of Colorado at Boulder, CO U.S.A.}

\author{Stephanie Swartz}
\affiliation{Time and Frequency Division, National Institute of Standards and Technology, Boulder, CO U.S.A.}

\author{Jens Biegert}
\affiliation{ICFO - Institut de Ciencies Fotoniques, The Barcelona Institute of Science and Technology, 08860 Castelldefels (Barcelona), Spain}
\affiliation{ICREA, Pg. Lluís Companys 23, 08010 Barcelona, Spain}

\author{Gregory B. Rieker}
\affiliation{Department of Mechanical Engineering, University of Colorado at Boulder, CO U.S.A.}

\author{Rohith Reddy}
\affiliation{Department of Electrical and Computer Engineering, University of Houston, TX U.S.A.}

\author{Scott A. Diddams}
\email{scott.diddams@colorado.edu}
 \affiliation{Department of Electrical, Computer and Energy Engineering, University of Colorado at Boulder, CO U.S.A.}
 \affiliation{Department of Physics, University of Colorado at Boulder, CO U.S.A.}

\date{\today}


\date{\today}

\begin{abstract}
Mid-infrared microscopy is an important tool for biological analyses, allowing a direct probe of molecular bonds in their low energy landscape. In addition to the label-free extraction of spectroscopic information, the application of broadband sources can provide a third dimension of chemical specificity. However, to enable widespread deployment, mid-infrared microscopy platforms need to be compact and robust while offering high speed, broad bandwidth and high signal-to-noise ratio (SNR). In this study, we experimentally showcase the integration of a broadband, high-repetition-rate dual-comb spectrometer (DCS) in the mid-infrared range with a scanning microscope. We employ a set of 1-GHz mid-infrared frequency combs, demonstrating their capability for high-speed and broadband hyperspectral imaging of polymers and ovarian tissue. The system covers 1000 $\mathrm{cm^{-1}}$ at \mbox{$\mathrm{\nu_c=2941 \; cm^{-1}}$} with 12.86 kHz spectra acquisition rate and 5 $\mathrm{\mu m}$ spatial resolution. Taken together, our experiments and analysis elucidate the trade-off between bandwidth and speed in DCS as it relates to microscopy. This provides a roadmap for the future advancement and application of high-repetition-rate DCS hyperspectral imaging.
\end{abstract}


\maketitle 


\section{Introduction}
Vibrational spectroscopy is a cornerstone technique for molecular characterization that provides detailed insights into molecular structures, interactions, and dynamics. Essentially, it identifies the unique vibrational ``fingerprints" of molecules, revealing their specific characteristics and behaviors. While this method provides deep insights into the molecular makeup, spectroscopic imaging has enhanced its scope. Spectroscopic imaging marries the depth of vibrational spectroscopy with spatial context, presenting a combined image where one can pinpoint not just which molecules are present but also their exact locations. This enhanced perspective is invaluable across diverse fields, from understanding material properties to probing biological systems, facilitating a richer and more precise understanding of intricate molecular landscapes.

Within this context, vibrational spectroscopy using mid-infrared light ($\sim$3 - 12 $\mathrm{\mu m}$) proves beneficial in bio-imaging because it provides label-free chemical contrast and can non-invasively identify the biomolecular composition of samples \cite{bakerUsingFourierTransform2014}. Tissues, cells, and other biological entities can be analyzed without external dyes or markers, preserving the sample's native state. This is essential in clinical applications, where assessing the safety and efficacy of external contrast agents can be challenging. Spectroscopic imaging can complement conventional diagnostic techniques as the sample remains undisturbed. While spectroscopic imaging offers invaluable chemical insights, its slow signal acquisition has hindered its widespread adoption in biomedical and clinical settings. The trade-off between speed and bandwidth in recent systems \cite{zhangDepthresolvedMidinfraredPhotothermal2016,koleDiscreteFrequencyInfrared2012,yehFastInfraredChemical2015}  based on mid-infrared laser sources can compromise its competitiveness against label-based fluorescence microscopy, the principal bio-imaging technology.

In this work, we introduce and explore the capabilities of dual comb spectroscopy (DCS)\cite{coddingtonDualcombSpectroscopy2016} for mid-infrared hyperspectral microscopy.  Our work utilizes a set of recently developed 1-GHz mid-infrared frequency combs \cite{hoghooghiBroadband1GHzMidinfrared2022} to integrate a dual-comb spectrometer with a confocal microscope. We capitalize on the 1 GHz repetition rate of the combs for rapid data acquisition to capture a full spectrum across \mbox{2595 $\mathrm{cm^{-1}}$}--\mbox{3890 $\mathrm{cm^{-1}}$} every 78 $\mu$s (corresponding to a dual-comb repetition rate difference of  \mbox{$\Delta f_r=12.86 \; \text{kHz}$}). As such, the system is among the fastest performers in the class of spectrometers covering over \mbox{1000 $\mathrm{cm^{-1}}$} in the mid-infrared, while maintaining 1 GHz (0.03 cm$^{-1}$) spectral resolution.

\section{Landscape of Mid-infrared Hyperspectral Imaging}
\begin{figure}[!h]
    \centering
    \includegraphics[width=\linewidth]{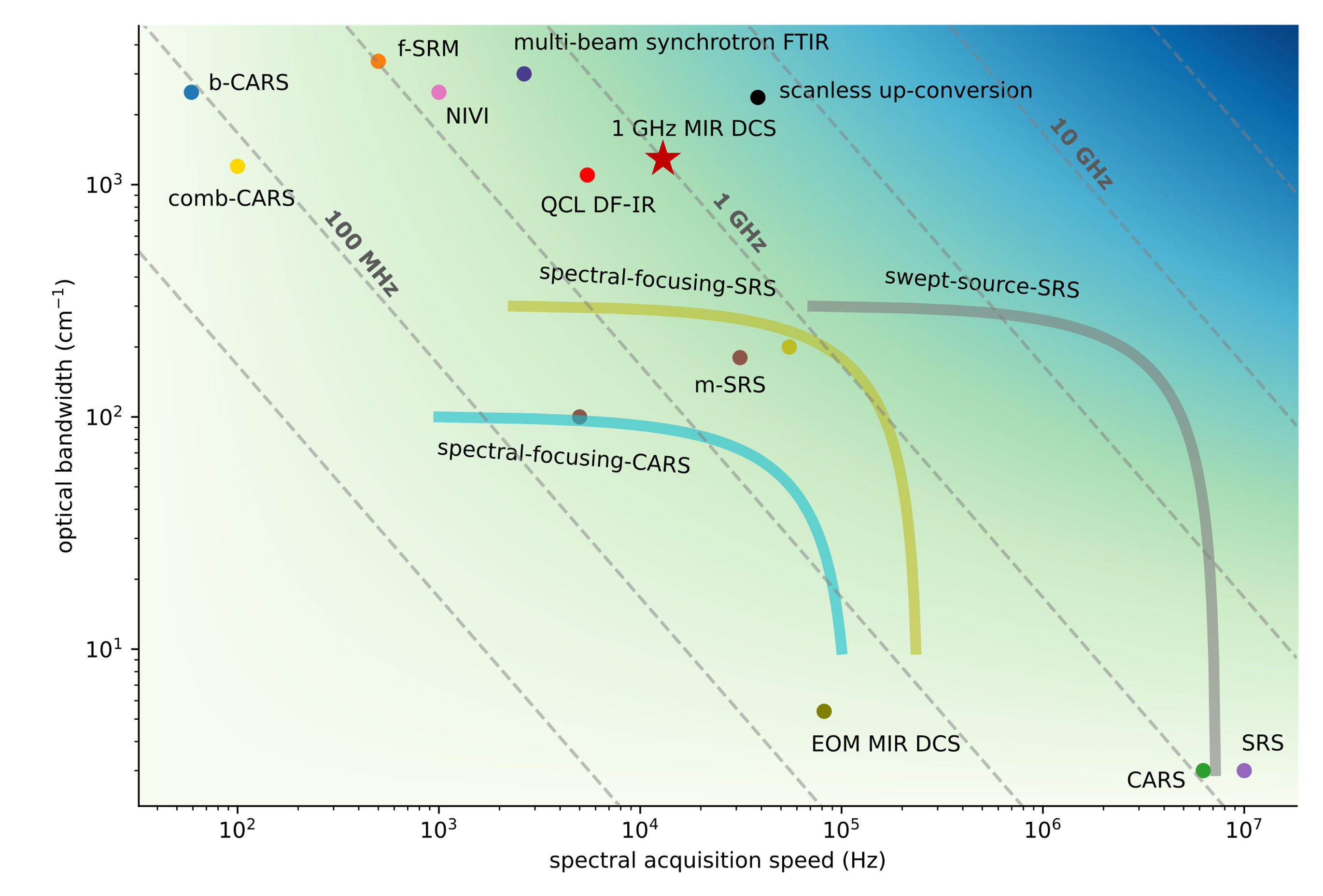}
    \caption{Performance map of mid-infrared hyperspectral imaging. As a guideline relevant to this work, the diagonal dashed lines show the $f_r^2/2$ trade-off inherent to DCS. Broadband CARS (b-CARS) \cite{keeSimpleApproachOnelaser2004}, femtosecond Stimulated Raman Microscopy (f-SRM) \cite{ploetzFemtosecondStimulatedRaman2007}, in-vivo video rate CARS \cite{evansChemicalImagingTissue2005}, in-vivo video rate SRS \cite{saarVideoRateMolecularImaging2010}, multiplexed SRS (m-SRS) \cite{fuQuantitativeChemicalImaging2012,liaoMicrosecondScaleVibrational2015}, nonlinear interferometric vibrational imaging (NIVI) \cite{chowdaryMolecularHistopathologySpectrally2010}, swept-source SRS \cite{ozekiHighspeedMolecularSpectral2012}, spectral-focusing SRS \cite{fuHyperspectralImagingStimulated2013, linMicrosecondFingerprintStimulated2021}, spectral-focusing CARS \cite{dinapoliHyperspectralDifferentialCARS2014}, comb-CARS \cite{ideguchiCoherentRamanSpectroimaging2013}, multi-beam synchrotron FTIR \cite{nasseHighresolutionFouriertransformInfrared2011}, electro-optic modulator comb MIR DCS \cite{ullahkhanDirectHyperspectralDualcomb2020}, QCL discrete frequency infrared imaging (DF-IR) \cite{yehFastInfraredChemical2015}, scanless mid-infrared up-conversion imaging \cite{zhaoHighspeedScanlessEntire2023}}
    \label{fig:bckgnd}
\end{figure}

The effort to push the barrier of spectral acquisition speed without compromising bandwidth is illustrated in \mbox{Fig. \ref{fig:bckgnd}}, which provides a visual overview of significant developments in hyperspectral vibrational imaging over the last two decades. Experiments are mapped onto the two important metrics of spectra acquisition speed, which captures the rate at which spectra are gathered, and optical bandwidth, which captures the breadth of chemical content that can be observed. The two variables are plotted against each other since a significant difficulty lies in achieving both metrics simultaneously. For a better one-to-one comparison, the acquisition speeds listed for experiments that use focal plane arrays have been normalized to that of an analogous point scanning experiment.


Coherent Raman spectro-imaging has achieved in-vivo video-rate speeds in the mid-infrared \cite{evansChemicalImagingTissue2005, saarVideoRateMolecularImaging2010} using well-established near-infrared femtosecond oscillators. Whereas initial demonstrations were over a narrow bandwidth (\mbox{$\sim$3 $\mathrm{cm^{-1}}$}), broad bandwidths at high acquisition speeds have been achieved using rapidly rotating polygonal mirror scanners \cite{tamamitsuUltrafastBroadbandFouriertransform2017, linMicrosecondFingerprintStimulated2021}. The high metrics are possible with the strong Raman absorption cross-sections around \mbox{2900 $\mathrm{cm^{-1}}$}. It is challenging, however, for Raman spectroscopy-based platforms to reach the same performance in the fingerprint region at longer wavelengths. An important target, therefore, is to achieve similar performance with direct mid-infrared illumination.

Fourier transform spectroscopy (FTS) and quantum cascade laser (QCL) based imaging are attractive due to their broad applicability across the mid to long wavelength infrared. The high absorption cross-sections alleviate the need for operation at powers close to sample-damage thresholds, a concern that is applicable to biological samples. In this category, FTS spectrometers coupled to broadband and bright sources such as synchrotron facilities have set the state of the art for the combination of spectral bandwidth and speed \cite{nasseHighresolutionFouriertransformInfrared2011}. However, a widely accessible imaging method would benefit from having a table top setup. To address this, QCL lasers are attractive due to their direct emission in the mid-infrared and small footprint, although their performance is best leveraged in narrowband applications. Tunable QCL packages consisting of multiple QCL chips combined into one device \cite{yehFastInfraredChemical2015,goyalActiveHyperspectralImaging2014,zimmerleiterQCLbasedMidinfraredHyperspectral2021} can reach broad spectral coverage, but further improvement is needed to reach noise figures comparable to platforms based on mode-locked lasers.



An alternative method to boost imaging speed in the mid-infrared is to employ up-conversion to shorter wavelengths in order to leverage low-cost near-infrared cameras, whose performance can significantly exceed mid-infrared focal plane arrays \cite{junaidVideorateMidinfraredHyperspectral2019,knezInfraredChemicalImaging2020,potmaRapidChemicallySelective2021}. A recent demonstration covered \mbox{$>$$1000 \; \mathrm{cm^{-1}}$} in eight seconds \cite{zhaoHighspeedScanlessEntire2023}. However, a potential drawback to this platform is a demand for very high pump pulse energies on the millijoule scale, requiring large regenerative amplifiers and a bulky apparatus.


Alongside scanless methods, the concept of a compact and deployable imaging system motivates parallel development of platforms seeded by well-developed fiber-integrated light sources. Fiber oscillators and amplifiers in the near infrared (for example, in the 1550 nm telecommunications band) are compact, robust and alignment free. However, their pulse energies typically fall in the 1 - 10 nanojoule range. This necessitates well-designed nonlinear optics to translate to the mid-infrared, as well as the development of techniques separate from existing scanless imaging.




In this work, we explore and apply high rep-rate dual-comb spectroscopy (DCS) to hyperspectral imaging. DCS has emerged as a powerful technique due to its combination of resolution, stability, and speed when compared to classical FTS \cite{coddingtonDualcombSpectroscopy2016}. In this modality, the interference of two frequency combs maps a Nyquist band from the optical domain down into the RF. One of the most important considerations in DCS is the direct trade-off between the repetition rate $f_r$ and the size of the optical Nyquist window created by the interference of the comb lines $\Delta \nu$:
\begin{align}
    \Delta \nu = \frac{f_r^2}{2 \Delta f_r}
    \label{eq:dcs}
\end{align}
where $\Delta f_r$ is the interferogram acquisition rate equal to the difference of the two laser rep-rates. The diagonal dashed lines in \mbox{Fig. \ref{fig:bckgnd}}., show the $f_r^2/2$ trade-off between resolvable bandwidth and acquisition speed in DCS for different $f_r$. Evidently, when the broad absorption features of large condensed phase molecules such as lipids and proteins allow for coarse resolution, the highest rep-rates exceeding 1 GHz are desired. 

As highlighted by the void in the upper right region of \mbox{Fig. \ref{fig:bckgnd}}, hyperspectral imaging that simultaneously affords 1000 cm$^{-1}$ coverage at megahertz acquisition rates is challenging to achieve. However, this presents an opportunity for scanned imaging with DCS platforms, provided the existence of broad bandwidth mid-infrared frequency combs with 10 GHz mode spacing. Generally speaking, this is a challenging regime for frequency combs, although such mode-spacings can be realized on EOM platforms  \cite{kowligyMidinfraredFrequencyCombs2020}, and in a few cases have even been used for imaging \cite{ullahkhanDirectHyperspectralDualcomb2020,khanSubGHzOpticalResolution2023}. Combs spanning bandwidths greater than 1000 cm$^{-1}$ at 10 GHz have been demonstrated in the near-infrared \cite{carlsonBroadbandElectroopticDualcomb2020}, but a  challenge remains in efficient and broad bandwidth down conversion to the mid-infrared\cite{kowligyMidinfraredFrequencyCombs2020}. Nonetheless, this is a promising direction to explore, and our present effort with 1 GHz mid-infrared dual comb imaging is a step towards even higher bandwidths and acquisition rates as given by Eq. \ref{eq:dcs}.







\section{Roadmap for Imaging Speed with DCS}
Whereas Fig. 1 illustrates the significant challenge of optical bandwidth and spectral acquisition rate, it does not fully capture the metric of imaging speed, which in the end is determined by the averaging time needed to reach sufficient SNR at each pixel. Consequently, it is useful to have an experimentally driven map of imaging speed in the relevant case of point scanning dual-comb microscopy, where the target SNR and frequency resolution sets the pixel dwell time.

\begin{figure}[!h]
    \centering
    \includegraphics[width=\linewidth]{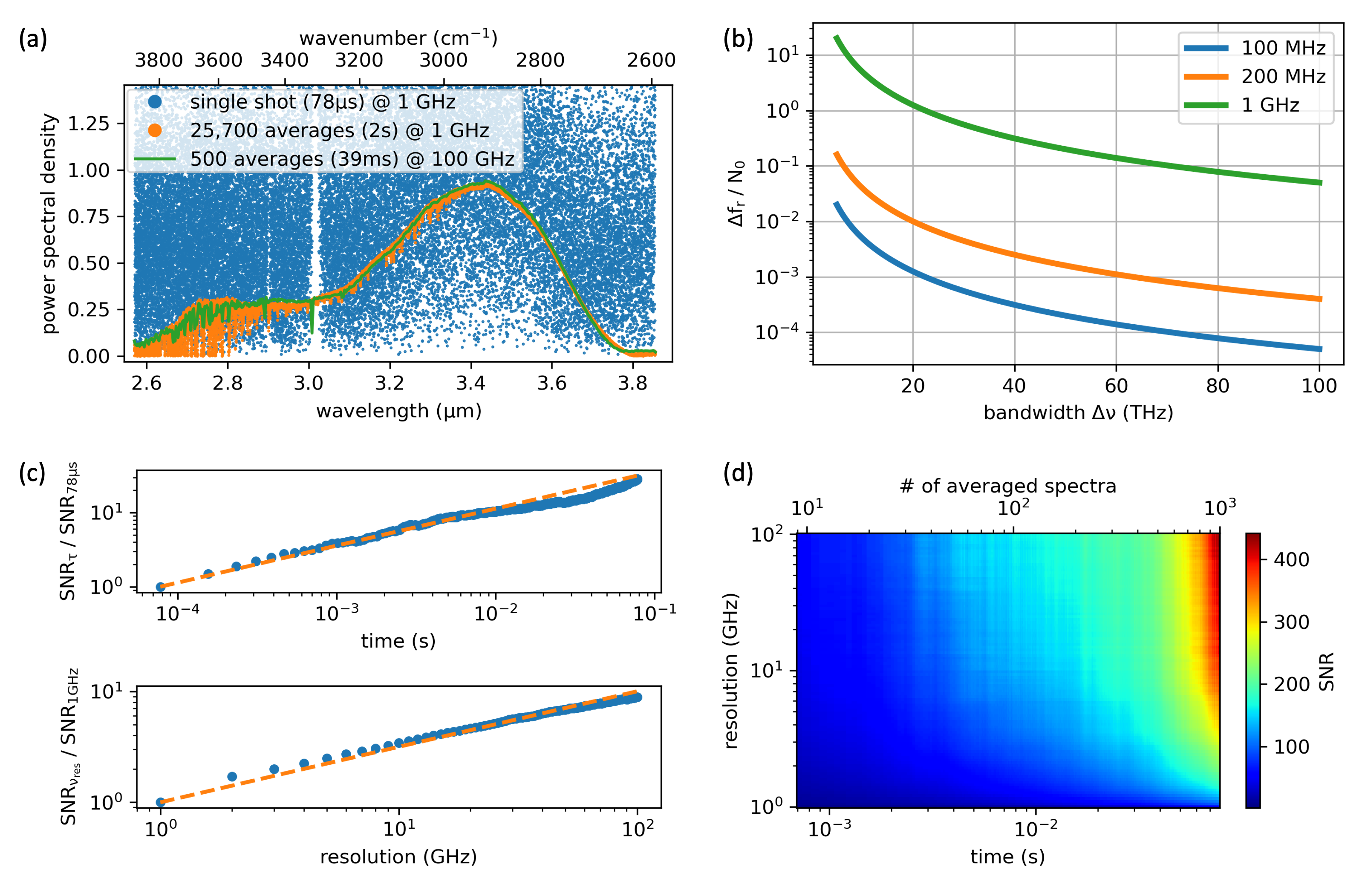}
    \caption{Summary of DCS imaging speed. (a) DCS spectra taken at different averaging times and frequency resolution / apodization windows. A single-shot spectrum at 1 GHz resolution can be averaged to high SNR in two seconds, or in 39 ms at 100 GHz resolution. The lines in the higher resolution spectrum are due to water absorption. (b) The spectra acquisition rate divided by the fundamental number of frequency bins, $N_0=\Delta \nu / f_r$, plotted against the size of the optical Nyquist window. Three curves are shown for different repetition rates. (c) The relative SNR improvement from signal averaging and apodization. The dashed orange curves mark the theoretical square root scaling. (d) The 2D SNR parameter space of signal averaging and apodization (combining the two plots in (c)) for the 1-GHz system.}
    \label{fig:snr_analysis}
\end{figure}

In DCS, the absorbance noise $\sigma$ scales with the frequency resolution and number of averaged spectra $N_{avg}$ according to \cite{newburySensitivityCoherentDualcomb2010}: 
\begin{align}
    \sigma \propto \frac{N}{\sqrt{N_{avg}}}\sqrt{\epsilon}
    \label{eq:snr}
\end{align}
where $N$ is the number of frequency bins and $\epsilon=\nu_{res}/f_{r}$ is referred to as the duty cycle, where $\nu_{res}$ is the frequency resolution and $f_r$ is the repetition rate. We note that $\epsilon\geq 1$, and only approaches unity for the case of mode-resolved spectroscopy with $\nu_{res}=f_r$, and otherwise increases with apodization according to $\epsilon\propto1/N$. The scaling rule for $\sigma$ is shown in \mbox{Fig. \ref{fig:snr_analysis}.(c).}, where it is observed to match the experimentally measured absorbance noise. Importantly, \mbox{Eq. \ref{eq:snr}} entails a square root scaling improvement in \mbox{$\mathrm{SNR}\propto 1/\sigma$} for both the number of averaged spectra and apodization of interferograms (decreasing $N$). The artificial resolution is penalized by the factor of $\sqrt{\epsilon}$, which softens the benefit of apodization to an effective $1/\sqrt{N}$ scaling. A directly proportional scaling improvement can be achieved, however, if one moves to a higher rep-rate source with larger mode-spacing.

This is illustrated in \mbox{Fig. \ref{fig:snr_analysis}.(a).} for the case of 1 GHz frequency combs. A baseline for 1 GHz DCS is that a 1000 $\mathrm{cm^{-1}}$ Nyquist window can be covered with \mbox{$\sim$17 kHz} spectra acquisition speed. In \mbox{Fig. \ref{fig:snr_analysis}.(a).}, a single-shot spectrum (78 $\mathrm{\mu s}$) at 1 GHz resolution has low signal to noise, but can be averaged to high SNR in two seconds ($>$25,000 spectra). However, \mbox{$\mathrm{SNR>200}$} can also be achieved in \mbox{$\sim$39 ms} at 500 averages if the interferograms are apodized to 100 GHz (\mbox{$\sim$3.33 $\mathrm{cm^{-1}}$}), which is a more appropriate sampling interval for the given absorption features. The SNR as a function of averaging time and apodized frequency resolution is shown in \mbox{Fig. \ref{fig:snr_analysis}.(c-d)}. The absorbance noise averages down according to $1/\sqrt{N_{avg}}$, and with coarser resolution resulting in a similar $1/\sqrt{N}$ noise reduction.

More importantly, this experimental analysis of bandwidth, SNR, frequency resolution, and imaging speed provides a roadmap for the integration of high rep-rate sources into the field of microscopy (\mbox{Fig. \ref{fig:bckgnd}}). The time, $\tau_{ifg}$, to acquire a single spectrum or interferogram decreases with the rep-rate according to $\tau_{ifg} \propto 1 / f_r^2$. However, an important additional consideration is the direct $\sigma \propto N$ improvement gained from the coarser resolution, compared to the $\sigma \propto \sqrt{N}$ improvement in the case of apodization. Practically speaking, for an equivalent optical nyquist band, the consideration of SNR adds an additional $f_r$ scaling to the well known $\Delta f_r \propto f_r^2$ (\mbox{Eq. \ref{eq:dcs}}), leading to an overall $f_r^3$ scaling with the mode-spacing (\mbox{Fig. \ref{fig:snr_analysis}.(b)}). As an example, in our previous work with 100 MHz mid-infrared DCS \cite{lindMidInfraredFrequencyComb2020,timmersMolecularFingerprintingBright2018} a comparable SNR was reached with \mbox{$\sim$ 1} minute averaging per pixel, whereas a potential 10 GHz DCS platform can reach the same SNR in 39 $\mathrm{\mu s}$ reducing the overall experiment time from the several hours in this experiment down to \mbox{$\sim$ 10} seconds.

\section{Experiment and Results}

\begin{figure}[!h]
    \centering
    \includegraphics[width=\linewidth]{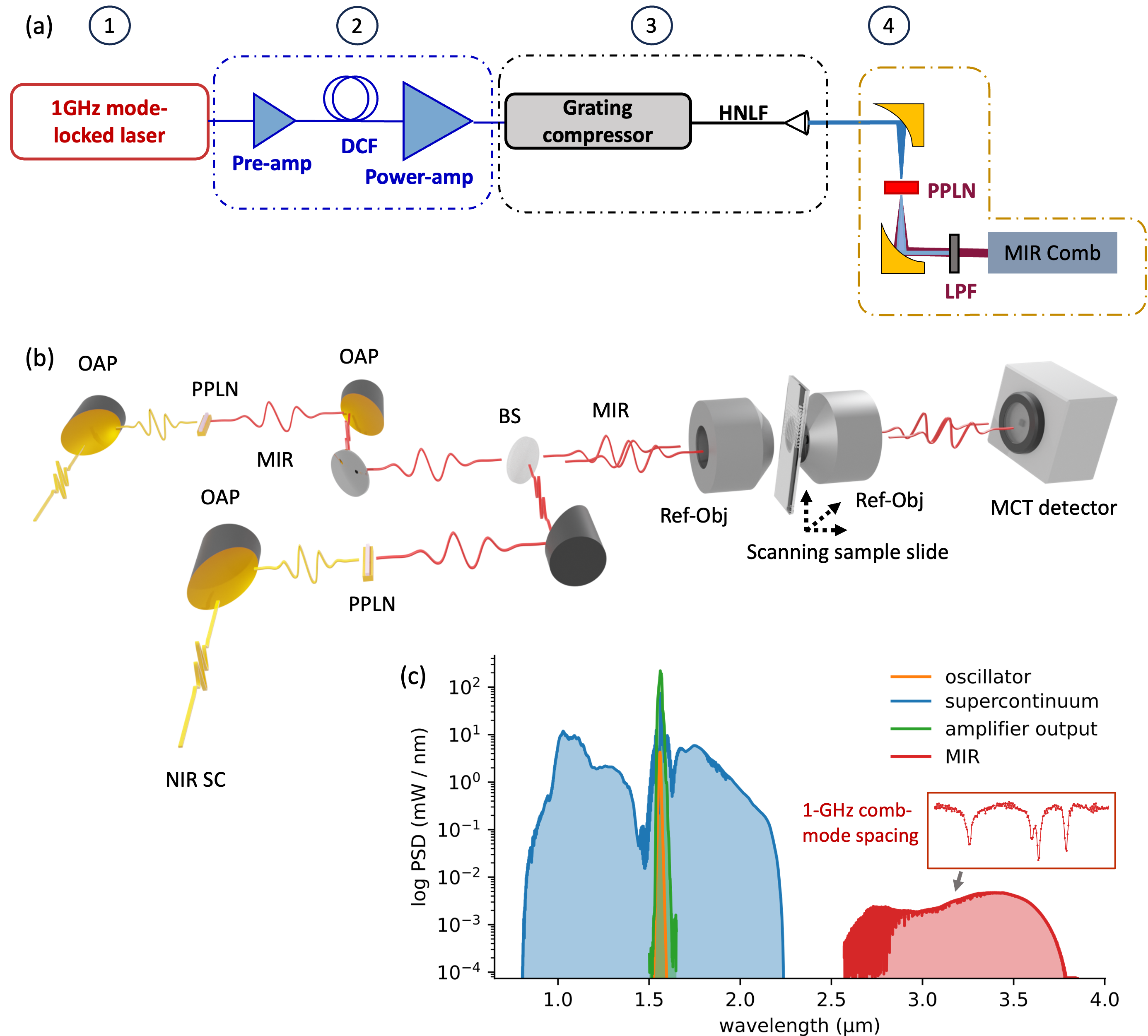}
    \caption{Experimental Setup. (a) Approach to generate mid-infrared frequency combs via soliton self-compression in highly nonliear fiber (HNLF) followed by intra-pulse difference frequency generation. (b) Two mid-infrared frequency combs generated through intra-pulse difference frequency generation in periodically-poled lithium niobate (PPLN) are passed collinearly through a confocal microscope. Hyperspectral images are collected by raster scanning the sample slide. The transmitted signal is collected and digitized in a high-speed MCT mid-infrared detector and FPGA. (c) Summary of near- and mid-infrared spectral envelopes.  The inset shows an example of spectroscopy of water vapor in ambient air with 1 GHz comb mode resolution}
    \label{fig:setup}
\end{figure}

With simplicity and long-term stability in mind, a single-branch intra-pulse difference frequency generation (DFG) design is used to generate frequency comb light in the mid-infrared\cite{lindMidInfraredFrequencyComb2020,kowligyMidinfraredFrequencyCombs2020,hoghooghiBroadband1GHzMidinfrared2022}. 
%
%
The architecture of a single mid-infrared comb source is illustrated in the upper part of \mbox{Fig. \ref{fig:setup}}. The 1 GHz mode-locked laser is amplified to 4 W and launched into highly nonlinear fiber (HNLF). Soliton self-compression in the anomalous dispersion HNLF results in octave-spanning, few cycle NIR pulses that efficiently drive the nonlinear intrapulse down conversion to the mid-infrared. Although coverage of the \mbox{6-13 $\mathrm{\mu m}$} wavelength region has been demonstrated for one laser system \cite{hoghooghiBroadband1GHzMidinfrared2022}, in this work widely available lithium niobate is used to cover the \mbox{3 - 5 $\mathrm{\mu m}$} wavelength window. In addition, the offset frequency and repetition rate of the frequency comb are controlled via servo loops. Further details  are given in earlier publications \cite{hoghooghiBroadband1GHzMidinfrared2022,hoghooghiGHzRepetitionRate2024}

Two 1-GHz mid-infrared frequency combs are generated in this manner and coupled into $\mathrm{InF_3}$ single-mode fiber for delivery to the experiment. The output beam is collimated with a two inch off-axis parabolic mirror, and a reflective confocal microscope with 0.58 NA is used to image the beam onto a glass slide (\mbox{$\sim$3.8 $\mathrm{\mu m}$} pixel size). A set of linear translation stages are used to raster scan the sample. The data is acquired via trigger, with the trigger spacing and scan speed set by the desired spatial sampling interval. The scan speed is limited by the interferogram acquisition time, which is fundamentally set by the repetition rate difference of the two combs, which is servo-controlled to be 12.86 kHz. The transmitted signal is focused onto a high-speed MCT detector, whose AC coupled port is digitized at 1 GS/s using an FPGA (GaGe RazorMax16). The data is streamed concurrently from the card memory into PC RAM for real-time analysis, and such that the card-memory does not limit the data volume. Owing to the fairly high 500 MHz Nyquist frequency, and the placement of all fiber amplifiers within the physical path of the servo control loops of the two frequency combs, over one thousand interferograms can be directly averaged before phase correction needs to be employed \cite{hebertSelfcorrectedChipbasedDualcomb2017,hebertSelfCorrectionLimitsDualComb2019}.

\begin{figure}[!h]
    \centering
    \includegraphics[width=\linewidth]{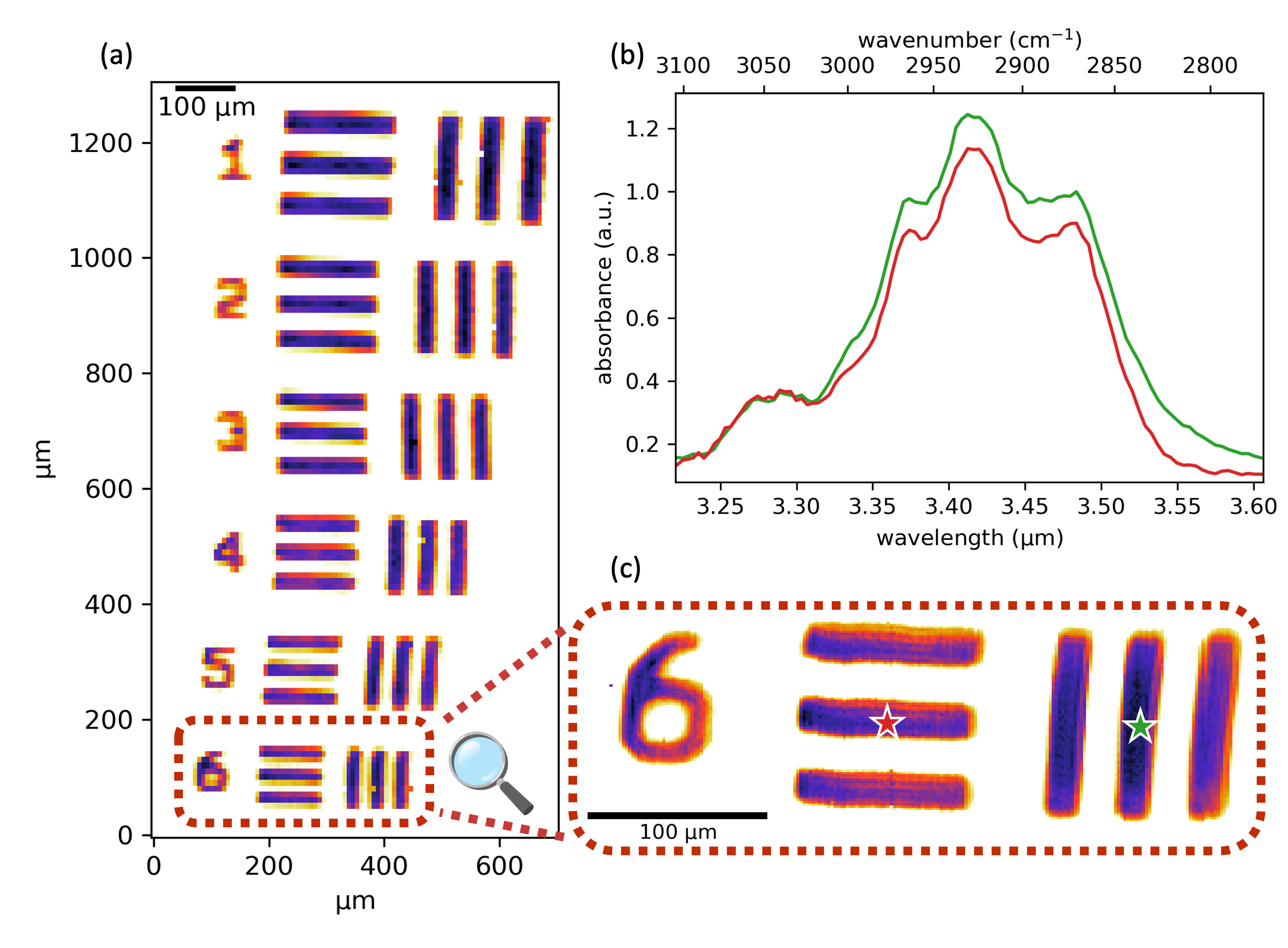}
    \caption{Dual-comb hyperspectral image of SU-8 USAF test pattern on Silicon. (a) False-color image generated from the integrated absorbance around \mbox{2930 $\mathrm{cm^{-1}}$} taken at \mbox{5 $\mathrm{\mu m}$} spatial sampling. (b) Absorbance curves taken at the red and green points (stars) of frame (c), which is a zoomed image taken at \mbox{1.2 $\mathrm{\mu m}$} spatial sampling.}
    \label{fig:su8}
\end{figure}

To verify the spatial resolution, hyperspectral images are taken of a USAF resolution target composed of SU-8 photoresist patterned onto a 500 $\mathrm{\mu m}$ thick Silicon wafer. Five hundred spectra (acquired in 39 ms) are averaged at each pixel and apodized to 100 GHz (\mbox{3.3 $\mathrm{cm^{-1}}$}). Point spectra shown in \mbox{Fig. \ref{fig:su8}.(b).} are taken at each pixel to generate the hypercube. The images are generated by integrating a \mbox{$\sim$63 $\mathrm{cm^{-1}}$} window around the peak absorption at \mbox{$\sim$2930 $\mathrm{cm^{-1}}$}. A spatial resolution of \mbox{$5 \; \mathrm{\mu m}$} is estimated from the line scans across the SU-8 bars.


\begin{figure}[!h]
    \centering
    \includegraphics[width=\linewidth]{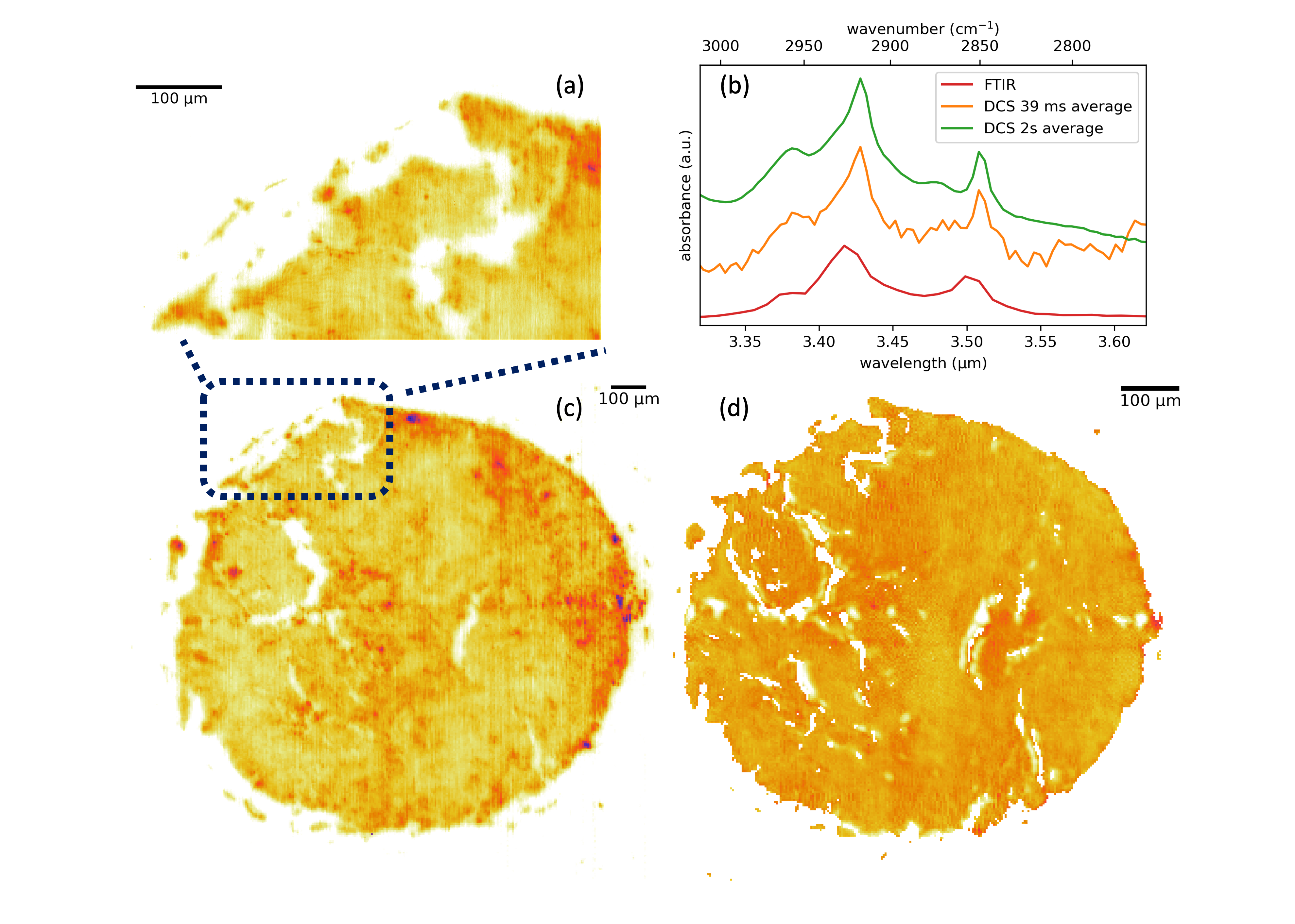}
    \caption{Hyperspectral image of ovarian cancer tissue. (a) Absorbance at \mbox{2920 $\mathrm{cm^{-1}}$} taken at \mbox{1.2 $\mathrm{\mu m}$} sampling. (b) Absorbance curves taken on cancer tissue with two second average in green, 39 ms average in orange, and with a commercial FTIR in red. Absorbance curves have been offset for clarity. (c) Zoom out of frame (a) to include absorbance image of the whole \mbox{1 mm} core at \mbox{2920 $\mathrm{cm^{-1}}$} taken with \mbox{5 $\mathrm{\mu m}$} spatial sampling. (d) Equivalent image to (c) but taken using a commercial FTIR microscope with a focal plane array.}
    \label{fig:bio}
\end{figure}

To illustrate the potential of our technique for biologically relevant samples, we image a cross-section of ovarian cancer tissue. Here, we also validate our DCS point scanning hyperspectral microscopy by comparison to data taken with a commercial FTIR microscope using a focal plane array.  For the DCS microscopy, five hundred spectra are again averaged at each pixel and apodized to \mbox{3.3 $\mathrm{cm^{-1}}$}. Point spectra such as the one shown by the orange curve in \mbox{Fig. \ref{fig:bio}.(b).} are collected at each pixel, with the two C-H anti-symmetric stretch bands visible at 2850 and \mbox{2920 $\mathrm{cm^{-1}}$}. A DCS spectrum taken with a two second averaging time (25,700 averages) is shown by the green curve, and a comparison spectrum taken using a commercial FTIR with \mbox{7.61 $\mathrm{cm^{-1}}$} frequency resolution is shown by the red curve. Apart from a broadening of the peak, good agreement is observed between the DCS and FTIR spectra. The FTIR image was taken prior to the removal of paraffin wax covering the sample, which accounts for the peak broadening when compared to the spectra taken using DCS where no paraffin was present. 

The images are generated by taking a slice through the hypercube at the peak of the \mbox{2920 $\mathrm{cm^{-1}}$} band. A coarse image shown in \mbox{Fig. \ref{fig:bio}.(c).} with \mbox{5 $\mathrm{\mu m}$} sampling is taken of the entire \mbox{1 mm} core. A zoom-in of the sample is shown in \mbox{Fig. \ref{fig:bio}.(a)}, taken at \mbox{1.2 $\mathrm{\mu m}$} sampling. Considering the mechanical scan rate, the dwell time at each \mbox{1.2 $\mathrm{\mu m}$} pixel is \mbox{$\sim 40$ ms}, which is approximately the Nyquist sampling limit of the microscope. The image shows generally good agreement with the corresponding image in \mbox{Fig. \ref{fig:bio}.(d).} taken using the FTIR microscope. We note that the dim vertical line scans in the DCS image are attributed to the limited \mbox{$\sim$1.5 $\mathrm{\mu m}$} repeatability of the translation stages (Thorlabs Z825B).


\section{Discussion and Conclusion}

In this paper, we demonstrate a high rep-rate and broadband DCS microscope covering over $1000 \; \mathrm{cm^{-1}}$ in the 3-5 $\mu$m region of the mid-infrared. The results are used to create a performance map for future DCS applications in microscopy. At a 1 GHz rep-rate, DCS brings the point scanning system's overall experiment time to comparable numbers as FTIR microscopes that employ focal plane arrays to increase their speed. In such a case, acquisition times of a few hours are needed for a 512x512 image after averaging at each pixel. The DCS point scanning system offers a frequency axis calibrated to an uncertainty of the measured rep-rate around $\mathrm{10^{-11}}$, and whose resolution can be chosen down to the comb mode-spacing. The DCS point scanning system also offers the potential for future advances, such as multi-dimension compressed sensing. Importantly, compression could be performed not only on spatial sampling, but also spectrally by employing a time-programmmable frequency comb to implement a real-time apodization, as recently show in other systems \cite{tourigny-planteApodizationDualcombSpectroscopy2020,kawaiCompressiveDualcombSpectroscopy2021,caldwellTimeprogrammableFrequencyComb2022}. From known work, compressively sampling the time domain interferogram could effectively push the system past the 10 GHz mode-spacing regime using digital locking electronics alone, and introduce accelerated imaging times reduced by factors over a hundred. We anticipate that further work in combining advanced combinations of platforms utilized by the DCS gas-sensing and imaging communities can realize a compact and robust solution to broad-band and high speed vibrational imaging. 

\section{Data Availability}
The data that support the findings of this study are available from the corresponding author upon reasonable request.


%
%

%


\bibliography{references}

\end{document}